# Conceptual Analogies Between Multi-Scale Feeding and Feedback Cycles in Supermassive Black Hole and Cancer Environments

Matteo Santoni[1]*, Francesco Tombesi[2,3,4,5], Alessia Cimadamore[6], Rodolfo Montironi[6] and Francesco Piva[7]

[1] Oncology Unit, Macerata Hospital, Macerata, Italy, [2] Physics Department, University of Rome "Tor Vergata", Rome, Italy, [3] Istituto Nazionale di Astrofisica, Astronomical Observatory of Rome, Monte Porzio Catone, Italy, [4] Department of Astronomy, University of Maryland Department of Astronomy, College Park, Maryland, MD, United States, [5] National Aeronautics and Space Administration/Goddard Space Flight Center, Greenbelt, MD, United States, [6] Section of Pathological Anatomy, Polytechnic University of the Marche Region, School of Medicine, United Hospitals, Ancona, Italy, [7] Department of Specialistic Clinical and Odontostomatological Sciences, Polytechnic University of Marche, Ancona, Italy

Adopting three physically-motivated scales ("micro" – "meso" – "macro", which refer to mpc – kpc – Mpc, respectively) is paramount for achieving a unified theory of multiphase active galactic nuclei feeding and feedback, and it represents a keystone for astrophysical simulations and observations in the upcoming years. In order to promote this multi-scale idea, we have decided to adopt an interdisciplinary approach, exploring the possible conceptual similarities between supermassive black hole feeding and feedback cycles and the dynamics occurring in human cancer microenvironment.

## INTRODUCTION

Several scientific problems show complexities that can be understood only through a multi-scale approach, in which apparently disjointed processes are linked together across multiple scales. One of such examples is the astrophysical study of supermassive black holes (SMBHs), which resides at the core of virtually every galaxy in the universe. Feeding of matter from galactic distances onto SMBHs is thought to be the physical process powering active galactic nuclei (AGN), which are observed up to extremely large distances due to their large luminosities and are supposed to impact the evolution of their host galaxies through mechanical feedback from winds and jets (1–3).

It is important to remind that the typical scale associated with a SMBH is in units of its Schwarzschild radius $rs = 2GM_{BH}/c^2$, where $G$ is the gravitational constant, $c$ is the speed of light, and $M_{BH}$ is the mass of the SMBH. For a value of the SMBH mass of $\simeq 10^{10}$ times that of our Sun, this corresponds to a scale in units of parsec of 1 mpc $= 10^{-3}pc \simeq 10^{14\circ} cm$. This scale is larger than the distance between our Earth and the Sun, but it is negligible when compared to the size of a typical galaxy that can span a radius of more than 10 kpc. Therefore, the study of AGN and their host galaxies is inherently a multi-scale problem.

It has been recently assessed that adopting three physically-motivated scales ("micro" – "meso –macro, which refer tompc – kpc – Mpc, respectively) is paramount for achieving a unified theory of multiphase AGN feeding and feedback, and it represents keystone for astrophysical simulations and observations in the upcoming years (4). In order to promote this multi-scale idea, we have decided to adopt an interdisciplinary approach, exploring the possible conceptual similarities between SMBH feeding and feedback cycles and the dynamics occurring in human cancer microenvironment.

To this end, analogously to the three major scales ("micro", "meso" and "macro") identified in the aforementioned astrophysical investigation, we defined three distinct scales within the tumor microenvironment: (1) micro ≡intratumor network; (2) meso≡ intercellular exchanges between tumor and immune cells; and (3) macro ≡extracellular vesicles-based communication between primary tumor and distant pre-metastatic niches. In these three contexts, cell-to-cell communication is mediated by exosomes, nano-sized vesicles (40-100 nm in diameter) enclosed by a lipid bilayer and released by cells under healthy or pathological conditions (5) (**Figure 1**). Exosomes contain a variety of biomolecules that include oncogenic proteins, signaling molecules, glycans, lipids, metabolites, RNA, and DNA (6).

## MICRO: EXOSOME-BASED INTRATUMOR COMMUNICATION

Studies focused on a variety of distinct cancer cells reported that tumor-derived exosomes can promote tumor cell proliferation. An autocrine induction of cellular proliferation was reported in chronic myeloid leukemia (7), gastric cancer (8), bladder cancer (9), glioblastoma (10) and melanoma (11). Furthermore, exosomes-based intratumor communication is also fundamental for the acquisition of migratory properties in cancer cells from the primary tumor. This phenomenon has been reported in both nasopharyngeal carcinoma (12) and prostate cancer (13), where exosomes released by primary tumor cells can increase the invasiveness and motility of other recipient malignant cells (**Figure 2**).

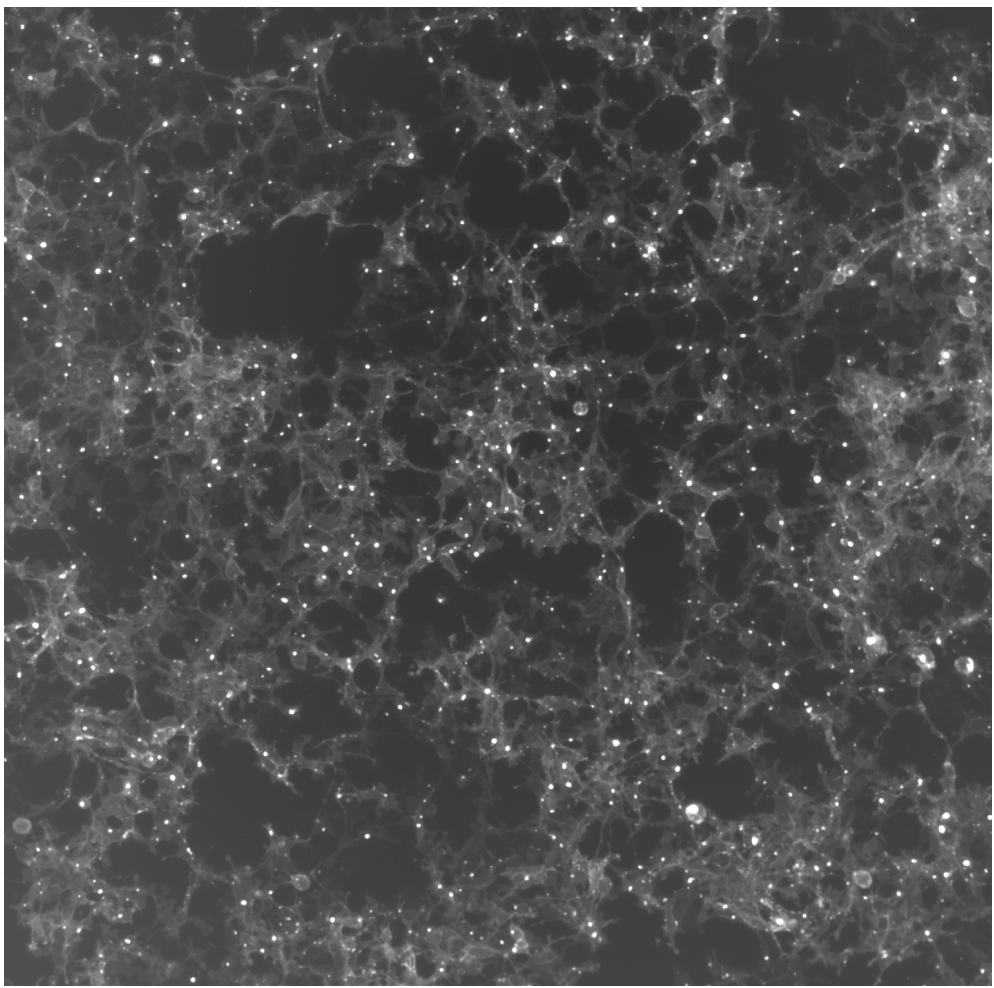

**FIGURE 1** | Fluorescence image of HEK293T cells in culture. The clear areas indicate a high concentration of exosomes. These cells were engineered to produce exosomes containing the GFP fluorescent protein.

A single cancer cell can produce and release in the blood from 2000 to 7000 exosomes (14). In 2018, Professor Avner Friedman and Professor Wenrui Hao (15) developed a mathematical model that resumes the production of tumor exosomes and the release of their content upon encountering cancer cells:

$$\frac{\partial Ec}{\partial t} - D_{Ec}\Delta E_c = \underbrace{\lambda_{Ec}C}_{production} - \underbrace{D_{Ec}\,E_c\left(\frac{C}{C+K_C}\right)}_{degradation}$$

where $E_c$ identifies the exosomes produced by cancer cell over time, $t$ = time, $D_{Ec}$ is the diffusion coefficient of $E_C$, $\lambda_{Ec}$ is the production rate of $E_C$, $C$ means Cancer cell density and $Kc$ indicates cancer cell saturation.

Exosomes released by tumor cells provide a paracrine signaling mechanism for cancer progression. Exosomes contain microRNAs (miRNAs), lipids and proteins that are cell type specific. The secretion and delivery of exosomal miRNAs are the basis for cancer cell-to-cell communication and contribute as signaling molecules to the creation of a tumor-promoting environment (16). The transfer of exosomal miRNAs can confer also acquired drug resistance by encoding proteins that can lead to chemoresistance in the recipient tumor cells (17).

miR-21 is one of the most studied (18) and enhances tumor growth when is released by exosomes through the encounter between $E_c$ and cancer cells, as expressed by the equation (15):

$$\frac{\partial\, miR21}{\partial t} - DmiR - 21\Delta_{miR21}$$

$$= \underbrace{\lambda_{miR21Ec}\,Ec\,\frac{C}{C+Kc}}_{production} - \underbrace{d_{miR21}\,miR-21}_{degradation}$$

where $D_{miR-21}$ is the diffusion coefficient of miR21once released by exosomes, estimated at 0.130 cm²/day (19), $\lambda_{miR21Ec}$ represents the production rate of miR-21 by $Ec$, $Kc$ indicates cancer cell saturation (estimated at 0.4 g/cm³) (20) and $d_{miR21}$the degradation rate of miR-21.

## MESO: EXOSOME-BASED DIALOGUE BETWEEN TUMOR AND IMMUNE CELLS

The ability to develop strategies to escape from host immune surveillance is one of the hallmarks of cancer (21). It has been shown that tumor-derived exosomes can down-regulate CD3ζ and Janus kinase 3 (JAK3) expression in primary activated T-cells, mediate the apoptosis of CD8+ T-cells and the conversion of CD4+ CD25- T-cell into CD4+CD25(hi)FOXP3+regulatory T-cells, (which express interleukin 10, transforming growth factor β1 and cytotoxic T-lymphocyte antigen 4 that effectively mediate suppression) (22) (**Figure 2**).

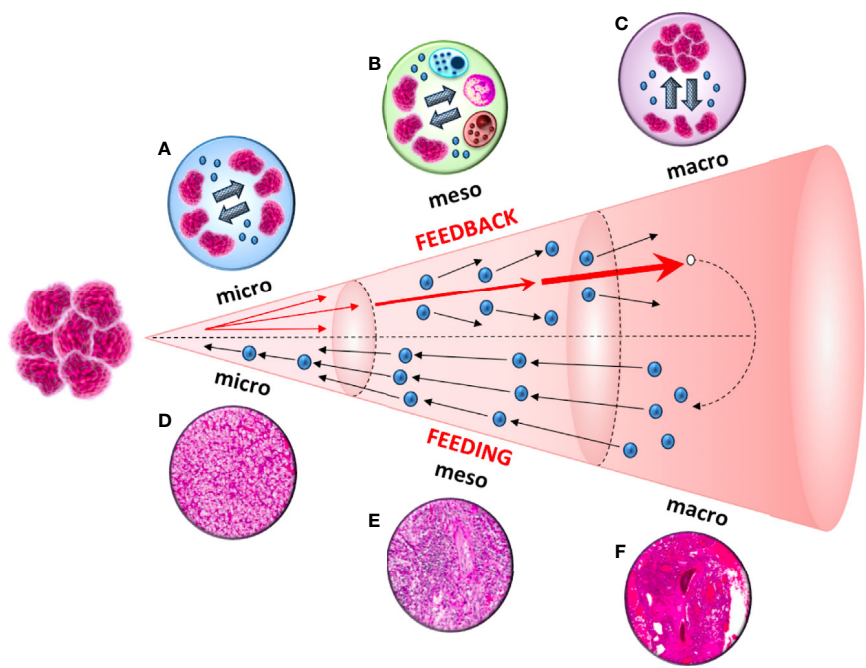

**FIGURE 2** | A diagram of the feeding and feedback cycles, showing the geometric increase of three orders of magnitude (micro, meso and macro) in tumor microenvironment. Tumor-derived exosomes can increase the invasiveness and motility of recipient cells. This exosome-mediated communication occurs at three different levels: between cancer cells in the primary tumor (micro), between cancer and immune cells (meso) and between cancer cells from the primary tumor to the metastatic sites (macro).

$E_c$ released by tumor cells release their content through the encounter with a variety of immune cells including Th1 cells ($T_4$), CD8+ cells ($T_8$), regulatory T cells ($T_r$) and dendritic cells ($D$):

$$\underbrace{\lambda_{Ec}\,C}_{production} - \underbrace{d_{Ec}E_c\left(\frac{T_4}{T_4 + K_{T4}} + \frac{T_8}{T_8 + K_{T8}} + \frac{T_r}{T_r + K_{Tr}} + \frac{D}{D + K_D}\right)}_{degradation}$$

where $T_4$, $T_8$ $T_r$ and $D$ express different immune cell density, while $K$ indicate cell saturation estimated at $2 \times 10^{-3}$ g/ml for $T_4$ and $T_8$, $5 \times 10^{-4}$ g/ml for $T_r$ (20, 23) and $4 \times 10^{-6}$ g/cm$^3$ for $D$ (24–26). The complex interactions of $T_8$, $D$ and tumor cells ($T$) within the tumor has been described by Depillis and colleagues (27), who extended the model previously elaborated by Ludewig *et al.* (28) by adding a tumor compartment. In particular, they took into account tumor-immune system parameters such as immune cell trafficking rates to and from the tumor, effector cell deactivation rates by tumor cells, effector cell death rates, intrinsic tumor growth rates, and tumor cell kill rates by effector cells (27). The interactions of $T_8$, $D$ and $T$ were described by:

$$\frac{d}{dt}E_{tumor}^a = \mu_{BTE}(T)E_{blood}^a - a_{EaT}E_{tumor}^a - cE_{tumor}^a T$$
$$\underbrace{}_{\text{Interactions between } CD8^+ \text{ and tumor cells}}$$

where $E_{tumor}^a$ is the number of CD8$^+$ cells within the tumor, $T$ is the number of tumor cells, $\mu_{BTE}$ is the T-dependent rate at which effector cells enter the tumor compartment from the blood, $E_{blood}^a$ is the number of CD8$^+$ cells in the blood, $a_{EaT}$ is the death rate of activated CD8$^+$ in the tumor [estimated at 0.462/day (29)], $c$ is the rate at which activated CD8$^+$ are inactivated by T [estimated at $9.42 \times 10^{-12}$ cells x day (27)];

$$\frac{d}{dt}T = rT\left(1 - \frac{T}{k}\right) - DT \cdots\cdots with \quad D = d\frac{\left(\frac{E_{tumor}^a}{T}\right)^l}{s + \left(\frac{E_{tumor}^a}{T}\right)^l}$$
$$\underbrace{}_{\text{Tumor growth and lysis by } CD8^+}$$

where $r$ is the tumor growth rate [estimated at 0.3954/day (30)], $k$ is the carrying capacity of tumor [estimated at $1.0 \times 10^9$ cells (30)] $s$ is the value of $\frac{E_a^a}{T}$ necessary for half-maximal activated CD8$^+$ toxicity and $l$ indicates the immune strength scaling exponent;

$$\frac{d}{dt}D_{tumor} = \frac{mT}{q + T} - (\mu_{TB} + a_D)D_{tumor}$$
$$\underbrace{}_{\text{Interactions between tumor and dendritic cells}}$$

where $D_{tumor}$ is the number of tumor-infiltrating dendritic cells, $m$ is the maximum recruitment rate of DCs to tumor site [estimated at $2.4388 \times 10^4$ cells/day (27)], $q$ is the value of $T$ necessary for half-maximal DC recruitment [estimated at 100 cells (27)], $\mu_{TB}$ is the rate of dendritic cell transfer from tumor to blood [estimated at 0.0011/day (27)], and $a_D$ is the natural death rate of dendritic cells [0.2310/day (28)].

# MACRO: EXOSOME-BASED COMMUNICATION BETWEEN PRIMARY TUMOR AND METASTASES

As reported above, tumor-derived exosomes promotes the invasiveness and motility of tumor cells from the primary site to metastatic niches (12, 13) (**Figure 2**). By analyzing different cell lines, Yu et al. (31) observed 79 proteins that were differently expressed in exosomes between more and less metastatic tumors, and these proteins were implicated in cell adhesion, invasion, growth, metabolism and metastasis (31). Tumor-derived exosomes are crucial for the formation of pre-metastatic niches (32), which are necessary for the creation of a suitable environment for circulating tumor cells (CTCs) colonization and growing within a secondary site (33). The choice of metastatic sites is orchestrated by the primary tumor through secreted factors, with exosomes representing the main extravescicular population that mediates long-range signaling during metastasis (34–36). Exosome proteomics revealed distinct integrin expression patterns that could be used to predict organ-specific metastasis (37). For example, exosomal integrins α6β4 and α6β1 were linked to lung metastases, while αvβ5 was associated to liver metastases (37), supporting the organotropism of tumor-derived exosomes.

If we assume that the production of tumor exosomes $E_c$ and the release of their content upon encountering cancer cells can be described by:

$$\frac{\partial Ec}{\partial t} - D_{Ec}\Delta c = \underbrace{\lambda_{Ec}C}_{production} - \underbrace{d_{Ec}Ec\left(\frac{C}{C + K_C}\right)}_{degradation} \quad [12]$$

We can derive that the diffusion of tumor-derived exosomes from primary tumor to pre-metastatic niches ($D_{Ec}\Delta c$) can be described by:

$$D_{Ec}\,\Delta c = \frac{\partial Ec}{\partial t} + d_{Ec}Ec\left(\frac{C}{C + K_C}\right) - \lambda_{Ec}C$$
$$\underbrace{}_{\text{Diffusion of exosomes from primary tumor to metastatic cells}}$$

Where $\lambda_{Ec}$ is the production rate of $E_c$ and $Kc$ indicates cancer cell saturation.

The diffusion of tumor-derived exosomes ($D_{Ec}$) from the primary tumor the pre-metastatic niche can be estimated through considering the average diameters of exosomes (70 nm) at$1.23 \times 10^{-4}$ cm$^2$/day (15). The release of $E_c$ promotes the initial phases of tumor invasion and metastatization, described by the model we adapted from that proposed by Lai and Friedman (38):

$$\frac{\partial C}{\partial t} - D_c\Delta C - \chi\nabla\,(C\,\nabla\,C) = \underbrace{(\lambda_{C1}\frac{SP_1}{K_{SP1} + SP_1}}_{Tumor\ invasion} + \underbrace{\lambda_{C2}\frac{SP_2}{K_{SP2} + SP_2}}_{Tumor\ growth} C\left(1 - \frac{C}{C_M}\right) - \underbrace{(d_pC + d_cC)}_{Tumor\ apoptosis}$$

where $D_c$ is the diffusion coefficient of cancer cells, $\chi \nabla (C \nabla C)$ indicates the migration of cancer cells, $\lambda_C$ the growth rate due respectively to driver signalling pathways 1 ($SP_1$) and 2 ($SP_2$), $C_M$ the carrying capacity of tumor cellsand $d_D C$ and $d_c C$ account, respectively, for the rate of damage-induced or natural apoptotic cancer cells.

## CONCLUSIONS

Analogies between multi-scale feeding and feedback cycles in supermassive black hole and cancer environments can allow the crossover of mathematical models between these two fields. Nevertheless, our model should be validated. The recent advances in tissue biology on several spatial scales including multi-cellular, single cell and the sub-cellular levels will expand our comprehension of the mechanisms of exosome-mediated communication and will allow us to measure the distances needed for cell-to-cell interactions. In this context, Nanostring technology can be employed to investigate tumor spatial and time heterogeneity. The potential consequences of the creation and validation of this model on the exosome-mediated communication between tumor cells and immune cells will be both clinical and therapeutical. Indeed, modelling exosome-mediated communication may allow to predict the time to metastases, to avoid tumor-induced immunosuppression and to develop exosomes for targeted delivery drug.

A straight cooperation and merging of such apparently distant disciplines may result a turning point to exceed the current limits of astrophysical and oncological modelling procedures in future years.

## AUTHOR CONTRIBUTIONS

MS: conceptualization and writing. FT: conceptualization and writing. AC: data collection. RM: supervision. FP: data collection and supervision. All authors contributed to the article and approved the submitted version.

## REFERENCES

1. Fabian AC. Observational Evidence of Active Galactic Nuclei Feedback Vol. 50. ARA&A (2012). p. 455. doi: 10.1146/annurev-astro-081811-125521

2. King A, Pounds K. Powerful Outflows and Feedback From Active Galactic Nuclei Vol. 53. ARA&A (2015). p. 115. doi: 10.1146/annurev-astro-082214-122316

3. Tombesi F, Melendez M, Veilleux S, Reeves JN, Gonzales-Alfonso E, Reynolds CS. Wind From the Black-Hole Accretion Disk Driving a Molecular Outflow in an Active Galaxy. Nature (2015) 519:436. doi: 10.1038/nature14261

4. Gaspari M, Tombesi F, Cappi M. Linking Macro, Meso, and Micro Scales in Multiphase Agn Feeding and Feedback. Nat Astronomy (2020) 4:10. doi: 10.1038/s41550-019-0970-1

5. Lin J, Li J, Huang B, Liu J, Chen X, Chen XM, et al. Exosomes: Novel Biomarkers for Clinical Diagnosis. Sci World J (2015) 2015:657086. doi: 10.1155/2015/657086

6. Mathieu M, Martin-Jaular L, Lavieu G, Théry C. Specificities of Secretion and Uptake of Exosomes and Other Extracellular Vesicles for Cell-Tocell Communication. Nat Cell Biol (2019) 21:9–17. doi: 10.1038/s41556-018-0250-9

7. Raimondo S, Saieva L, Corrado C, Fontana S, Flugy A, Rizzo A, et al. Chronic Myeloid Leukemia-Derived Exosomes Promote Tumor Growth Through an Autocrine Mechanism. Cell Commun Signal (2015) 13:8. doi: 10.1186/s12964-015-0086-x

8. Qu JL, Qu XJ, Zhao MF, Teng YE, Zhang Y, Hou KZ, et al. Gastric Cancer Exosomes Promote Tumour Cell Proliferation Through PI3K/Akt and MAPK/ERK Activation. Dig Liver Dis (2009) 41:875–80. doi: 10.1016/j.dld.2009.04.006

9. Yang L, Wu XH, Wang D, Luo CL, Chen LX. Bladder Cancer Cell-Derived Exosomes Inhibit Tumor Cell Apoptosis and Induce Cell Proliferation In Vitro. Mol Med Rep (2013) 8:1272–8. doi: 10.3892/mmr.2013.1634

10. Skog J, Würdinger T, van Rijn S, Meijer DH, Gainche L, Sena-Esteves M, et al. Glioblastoma Microvesicles Transport RNA and Proteins That Promote Tumour Growth and Provide Diagnostic Biomarkers. Nat Cell Biol (2008) 10:1470–6. doi: 10.1038/ncb1800

11. Matsumoto A, Takahashi Y, Nishikawa M, Sano K, Morishita M, Charoenviriyakul C, et al. Accelerated Growth of B16BL6 Tumor in Mice Through Efficient Uptake of Their Own Exosomes by B16BL6 Cells. Cancer Sci (2017) 108:1803–10. doi: 10.1111/cas.13310

12. You Y, Shan Y, Chen J, Yue H, You B, Shi X, et al. Matrix Metalloproteinase 13-Containing Exosomes Promote Nasopharyngeal Carcinoma Metastasis. Cancer Sci (2015) 106:1669–77. doi: 10.1111/cas.12818

13. Ramteke A, Ting H, Agarwal C, Mateen S, Somasagara R, Hussain A, et al. Exosomes Secreted Under Hypoxia Enhance Invasiveness and Stemness of Prostate Cancer Cells by Targeting Adherens Junction Molecules. Mol Carcinog (2015) 54:554–65. doi: 10.1002/mc.22124

14. Balaj L, Lessard R, Dai L, Cho YJ, Pomeroy SL, Breakefield XO, et al. Tumour Microvesicles Contain Retrotransposon Elements and Amplified Oncogene Sequences. Nat Commun (2011) 2:180. doi: 10.1038/ncomms1180

15. Friedman A, Hao W. The Role of Exosomes in Pancreatic Cancer Microenvironment. Bull Math Biol (2018) 80:1111–33. doi: 10.1007/s11538-017-0254-9

16. Falcone G, Felsani A, D'Agnano I. Signaling by exosomalmicroRNAs in Cancer. J Exp Clin Cancer Res (2015) 34:32. doi: 10.1186/s13046-015-0148-3

17. Qin X, Yu S, Zhou L, Shi M, Hu Y, Xu X, et al. Cisplatin-Resistant Lung Cancer Cell-Derived Exosomes Increase Cisplatin Resistance of Recipient Cells in Exosomal miR-100-5p-dependent Manner. Int J Nanomed (2017) 12:3721–33. doi: 10.2147/IJN.S131516

18. Sicard F, Gayral M, Lulka H, Buscail L, Cordelier P. Targeting miR-21 for the Therapy of Pancreatic Cancer. Mol Ther (2013) 21:986–94. doi: 10.1038/mt.2013.35

19. Bader AG, Brown D, Stoudemire J, Lammers P. Developing Therapeutic microRNAs for Cancer. Gene Ther (2011) 18:1121–6. doi: 10.1038/gt.2011.79

20. Cosio MG, Majo J, Cosio MG. Inflammation of the Airways and Lung Parenchyma in COPD: Role of T Cells. Chest (2002) 121:160S–5S. doi: 10.1378/chest.121.5_suppl.160S

21. Hanahan D, Weinberg RA. Hallmarks of Cancer: The Next Generation. Cell (2011) 144:646–74. doi: 10.1016/j.cell.2011.02.013

22. Whiteside TL. Immune Modulation of T-cell and NK (Natural Killer) Cell Activities by TEXs (Tumour-Derived Exosomes). Biochem Soc Trans (2013) 41:245–51. doi: 10.1042/BST20120265

23. Purwar R, Campbell J, Murphy G, Richards WG, Clark RA, Kupper TS. Resident Memory T Cells (T(RM)) are Abundant in Human Lung: Diversity, Function, and Antigen Specificity. PloS One (2011) 6:e16245. doi: 10.1371/journal.pone.0016245

24. Coventry BJ, Lee PL, Gibbs D, Hart DN. Dendritic Cell Density and Activation Status in Human Breast Cancer: CD1a, Cmrf-44, CMRF-56 and CD-83 Expression. Br J Cancer (2002) 86:546–51. doi: 10.1038/sj.bjc.6600132

25. Eden U, Fagerholm P, Danyali R, Lagali N. Pathologic Epithelial and Anterior Corneal Nerve Morphologyin Early-Stage Congenital Aniridickeratopathy. Ophthalmology (2012) 119:1803–10. doi: 10.1016/j.ophtha.2012.02.043

26. Troy AJ, Summers KL, Davidson PJ, Atkinson CH, Hart DN. Minimal Recruitment and Activation of Dendritic Cells Within Renal Cell Carcinoma. Clin Cancer Res (1998) 4:585–93.

27. Depillis L, Gallegos A, Radunskaya A. A Model of Dendritic Cell Therapy for Melanoma. *Front Oncol* (2013) 3:56. doi: 10.3389/fonc.2013.00056

28. Ludewig BB, Krebs P, Junt T, Metters H, Ford NJ, Anderson RM, et al. Determining Control Parameters for Dendritic Cell-Cytotoxic T Lymphocyte Interaction. *Eur J Immunol* (2004) 34:2407–18. doi: 10.1002/eji.200425085

29. dePillis LG, Gu W, Radunskaya AE. Mixed Immunotherapy and Chemotherapy of Tumors: Modeling, Applications and Biological Nterpretations. *J Theor Biol* (2006) 38:841–62. doi: 10.1016/j.jtbi.2005.06.037

30. Lee TH, Cho YH, Lee MG. Larger Numbers of Immature Dendritic Cells Augment an Anti-Tumor Effect Against Established Murine Melanoma Cells. *Biotechnol Lett* (2007) 29:351–7. doi: 10.1007/s10529-006-9260-y

31. Yu Z, Zhao S, Ren L, Wang L, Chen Z, Hoffman RM, et al. Pancreatic Cancer-Derived Exosomes Promote Tumor Metastasis and Liver Pre-Metastatic Niche Formation. *Oncotarget* (2017) 8:63461. doi: 10.18632/oncotarget.18831

32. Plebanek MP, Angeloni NL, Vinokour E, Li J, Henkin A, Martinez-Marin D, et al. Pre-metasta- Tic Cancer Exosomes Induce Immune Surveillance by Patrolling Monocytes At the Metastatic Niche. *Nat Commun* (2017) 8:1319. doi: 10.1038/s41467-017-01433-3

33. Peinado H, Zhang H, Matei IR, Costa-Silva B, Hoshino A, Rodrigues G, et al. Pre-metastatic Niches: Organ-Specific Homes for Metastases. *Nat Rev Cancer* (2017) 17:302–17. doi: 10.1038/nrc.2017.6

34. Janowska-Wieczorek A, Wysoczynski M, Kijowski J, Marquez-Curtis L, Machalinski B, Ratajczak J, et al. Microvesicles Derived From Activated Platelets Induce Metastasis and Angiogenesis in Lung Cancer. *Int J Cancer* (2005) 113:752–60. doi: 10.1002/ijc.20657

35. Emmanouilidi A, Paladin D, Greening DW, Falasca M. Oncogenic and non-Malignant Pancreatic Exosome Cargo Reveal Distinct Expression of Oncogenic and Prognostic Factors Involved in Tumor Invasion and Metastasis. *Proteomics* (2019) 19(8):e1800158. doi: 10.1002/pmic.201800158

36. Zhang H, Deng T, Liu R, Bai M, Zhou L, Wang X, et al. Exosome-Delivered EGFR Regulates Liver Microenviron- Ment to Promote Gastric Cancer Liver Metastasis. *Nat Commun* (2017) 8:15016. doi: 10.1038/ncomms15016

37. Hoshino A, Costa-Silva B, Shen TL, Rodrigues G, Hashimoto A, Tesic M, et al. Tumour Exosome Integrins Determine Organotropic Metastasis. *Nature* (2015) 527:329–35. doi: 10.1038/nature15756

38. Lai X, Friedman A. Exosomal microRNA Concentrations in Colorectal Cancer: A Mathematical Model. *J Theor Biol* (2017) 415:70–83. doi: 10.1016/j.jtbi.2016.12.006

**Conflict of Interest:** The authors declare that the research was conducted in the absence of any commercial or financial relationships that could be construed as a potential conflict of interest.